\documentclass[12pt,a4paper]{article}
\usepackage{amssymb,amsmath,graphicx,subcaption,hyperref,cite}
\usepackage{epstopdf}
\usepackage[utf8]{inputenc}
\usepackage{color}
\usepackage[english]{babel}
\begin{document}

\begin{center}
 {\Large\bf Critical Slowing Down along the Separatrix of Lotka-Volterra Model of Competition}
\end{center}
\vskip 1 cm
\begin{center} %
 Sauvik Chatterjee$^1$ and Muktish Acharyya$^{2,*}$
  
 \textit{Department of Physics, Presidency University,} 
  
 \textit{86/1 College Street, Kolkata-700073, India} 
 \vskip 0.2 cm
 {Email$^1$:chatterjeesauvik8@gmail.com}
  
 {Email$^2$:muktish.physics@presiuniv.ac.in}
\end{center}
\vspace {1.0 cm}
\vskip 1.5 cm 

\noindent {\large\bf Abstract:} The Lotka-Volterra model of competition has been studied by numerical simulations using the Runge-Kutta-Fehlberg algorithm. The stable fixed points, unstable fixed point, saddle node, basins of attraction, and the separatices are found. The transient behaviours associated with reaching the stable fixed point are studied systematically. It is observed that the time of reaching the stable fixed point in any one of the basins of attraction, depends strongly on the initial distance from the separatrix. As the initial point approached the separatrix, this time was found to diverge logarithmically. The divergence of the time, required to reach the stable fixed point, indicates the critical slowing down near the critical point in equilibrium phase transition. A metastable behaviour was also observed near the saddle fixed point before reaching the stable fixed point.


\vskip 2cm

\noindent $^*$ Corresponding author
\newpage

\section{Introduction}

The two species' competition for a limited resource is well described by the Lotka-Volterra model (LVM) in nonlinear dynamics\cite{strogatz}. A simple model could exhibit the separatrices and the basins of attraction. Let us mention some interesting research on LVM in the last few years. The extinction in the Lotka-Volterra model has been studied\cite{parker} to show the extinction time scale has a power law dependence on the population size. The extinction of neutrally stable LVM has also been investigated\cite{frey}. The deterministic approach results\cite{mobilia} coexistence of three species LVM in the stochastic cyclic LVM. The relations between stochastic LVM and the models of population genetics and game theory was established\cite{constable}. The stochastic LVM with interaction between predator and prey in a random environment has been investigated\cite{cai}. Competitive LVM has been investigated \cite{pigolotti} to study the spatial distribution (clustering) of the species for different kinds of interactions. In the Monte Carlo study\cite{provota} of lattice LVM, the clustering is found to show the fractal behaviours. An interesting phase transition was observed\cite{szabo} in the lattice LVM. The community of the nine toxicity/resistance types is found to undergo a phase transition as the uniform transmutation rates between the types decrease below a critical value. Above the critical value, all the nine types of strains coexist with equal frequency. The glassy phase was found\cite{altieri} in random LVM with demographic noise. In this kind of phase, the number of locally stable equilibria is an exponential function of the number of species.

The generalised LVM has been studied theoretically \cite{malcai} in the context of financial markets. The distribution of individual wealth or market values has been found to follow a power law. The Lotka-Volterra signals was observed\cite{marinakis} in ASEAN currency exchange rates. The efficient matrix technique has been developed \cite{izadi} to study the fractional LVM.

The effect of network structures on the population dynamics has been studied \cite{nagatani} in the diffusively coupled LVM stabilised by heterogeneous graphs. The asymptotic behaviour of a food-limited LVM (mutualism) with Markovian switching and Levy jumps has been studied \cite{liu}. The boundary effects on population dynamics in the stochastic lattice LVM have been studied\cite{heiba}. The system is observed to be divided into two regions, namely, the active coexistence and the predator extinction domain. The long-time behaviour of stochastic cooperative LVM with distributed delay has also been studied\cite{zuo}.

From the above mentioned studies on LVM, two things are clear to the reader. Firstly, the LVM is an ingredient of interesting modern research. Secondly, a wide variety of subjects are connected to LVM. However, it is also clear that the transient behaviour of the LVM has not been given much attention of the researchers. {\it What will be the transient behaviour near the separatices for the basins of attraction for the two species competition of LVM ?} 

In this short communication, we report the results of our studies
on the transient behaviours along the separatrix of LVM. The manuscript is organised as follows: the Lotka-Volterra model of competition is briefly described in the next section, the numerical results are given in section-3, the article ends with concluding remarks in section-4.

\section{Lotka-Volterra Model of Competition}
In this model, we consider two herbivorous species(Rabbits and Sheep) competing with each other over a limited amount of food supply. For simplicity, 
we consider\cite{strogatz} two different effects of this competition:

\begin{enumerate}
    \item Each species should grow upto their carrying capacity when the other one is absent. For our case, we assume the Rabbits having higher growth rate.
    
    \item When both the species face each other, then the struggle for food begins as the total food supply is limited. For this interaction, we safely assume that the species which are greater in number are heavier on the other. In spite of reduction of the growth rate of each species due to this interaction, we assume that this effect is heavier on the Rabbits.
\end{enumerate}

This model is expressed in the form of the two-variable coupled nonlinear deterministic differential equations given by:

\begin{align}
    \frac{dx}{dt} &= x(3-x-2y) \nonumber \\
    \frac{dy}{dt} &= y(2-x-y)
    \label{lv}
\end{align}

Where, $x(t)$ is the population of rabbits and $y(t)$ is the population of sheep.

Solving for $\Dot{x} = 0$ and $\Dot{y} = 0$ , we obtain $4$ fixed points: $(0,0)$,$(0,2)$,$(3,0)$ and $(1,1)$. $(0,0)$ is an \textit{unstable node}, $(0,2)$ and $(3,0)$ both are \textit{stable nodes} and $(1,1)$ is a \textit{saddle point}.

\vskip 0.5cm

\begin{figure*}[htpb]
 \centering
 (a)\includegraphics[width=0.62\columnwidth]{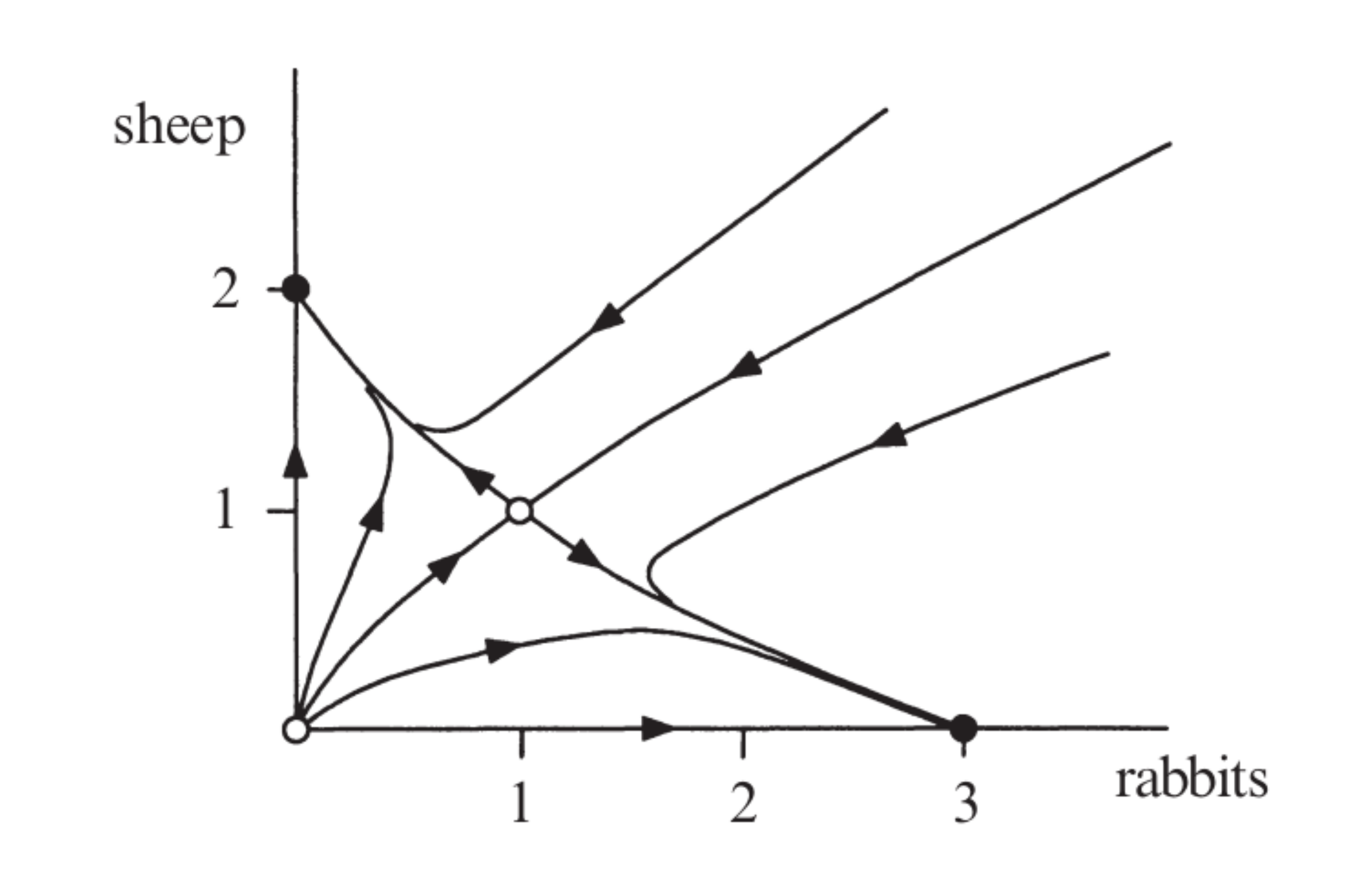}
 (b)\includegraphics[width=0.46\columnwidth]{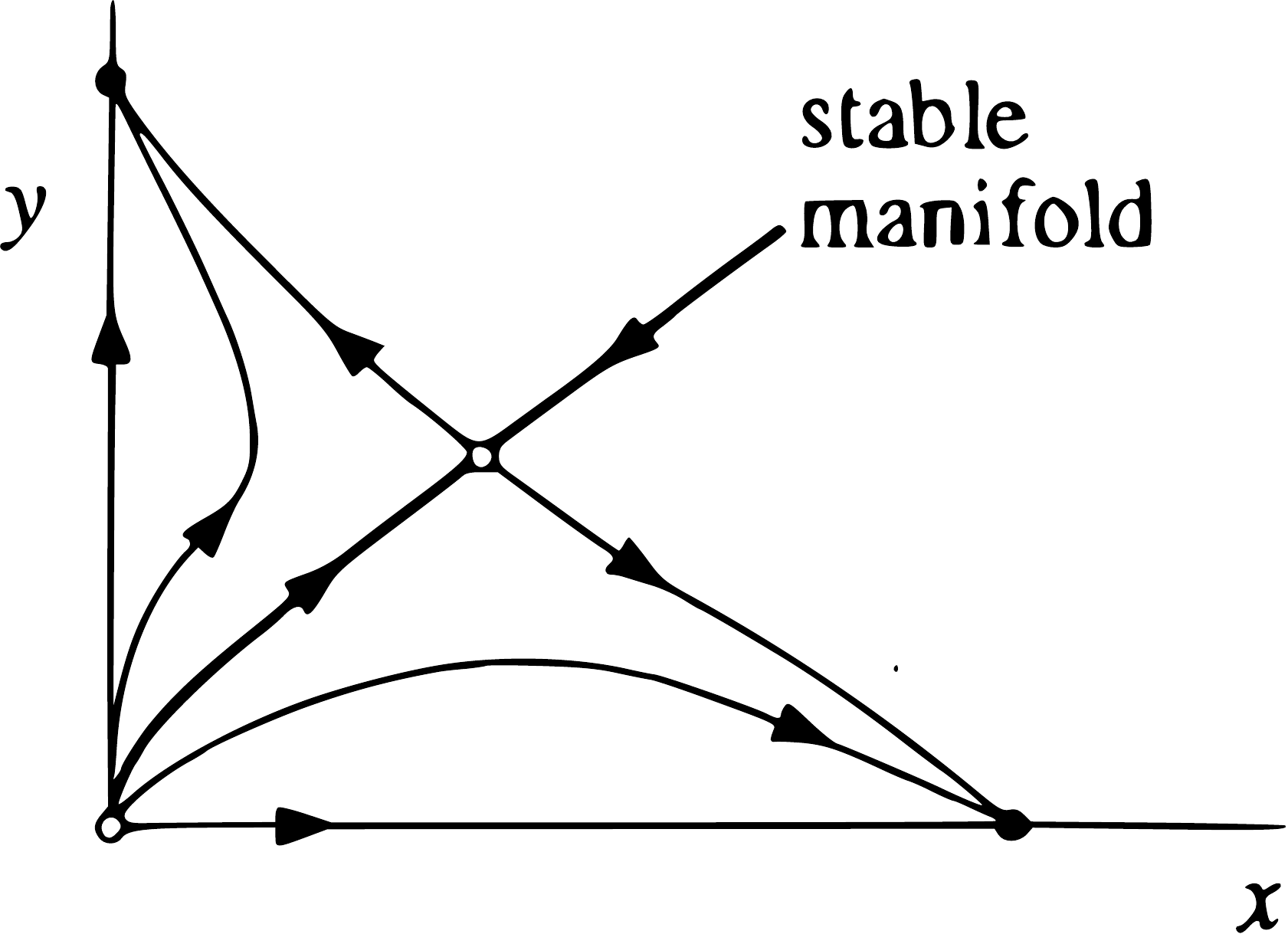}
 \caption{ Phase portrait of Lotka-Volterra Model : (a) The general phase portrait of rabbits and sheep along with the fixed points, (b) The stable manifold of saddle($1,1$) in the phase portrait(drawn with a heavy line). The plots are taken from the book {\it Nonlinear dynamics and Chaos}, S. H. Strogatz, (1994), Perseus Book publishing Limited, ISBN-0-201-54344-3
 }
 \label{fig:1}
\end{figure*}

From Figure-\ref{fig:1}, we can see that there can be three possibilities of a trajectory : (i) it can go to the stable node of x axis, (ii) it can go to the stable node of y axis, (iii) it dives into the saddle point which is actually at the \textit{stable manifold}.

From the phase portrait of Figure-\ref{fig:1}, we can also distinguish two different \textit{basins of attraction} (one for $(3,0)$ and one for $(0,2)$). The concept of basin of attraction is defined by setting an initial condition $z^*$ such that when $z(t) \to z^*$ ; $t \to \infty$ , where $z^*$ is an attracting fixed point. Depending on this basic mathematical concept, we proceed further in the next section.

\vskip 0.5cm

Figure-\ref{fig:1} signifies \textit{the principle of competitive exclusion}, which means that if two species compete over a limited source of food they can not simply coexist. We clearly observe that the trajectories below the stable manifold ultimately leads to the extinction of sheep and trajectories above the stable manifold leads to extinction of rabbits. There exists two separate basins of attraction which are separated by the separatrix (stable manifold). This existence of two different basins of attraction indicates the principle of competitive exclusion.

\section{Numerical Solution, Results and Discussion}

We analyse and discuss the results obtained from the basic mathematical concept of basin of attraction, described at the end of section 2. First we numerically solve both of equation \ref{lv}. Then we approach the separatrix through the stable manifold and discuss our results.

\subsection{Numerical Solution}

We have solved equation \ref{lv} using \textit{6th order Runge-Kutta-Fehlberg} technique to get the instantaneous value of population(both $x(t)$ and $y(t)$) as a function of time($t$).

\vskip 0.5 cm

This method of solving ordinary differential equation of the form $\frac{dz}{dt} = f(t,z(t)$ is based on the following 
numerical algorithm\cite{wheatley}:

\begin{equation}
    z(t+dt) = z(t) + \left(\frac{16 k_1}{135}+\frac{6656 k_3}{12825}+\frac{28561 k_4}{56430}- \frac{9 k_5}{50} + \frac{2 k_6}{55}\right)
    \label{rkf1}
\end{equation}

where

\begin{align}
    k_1 &= dt.f(t,z(t)) \nonumber \\
    k_2 &= dt.f\left(t+\frac{dt}{4},z+\frac{k_1}{4}\right) \nonumber \\
    k_3 &= dt.f\left(t+\frac{3dt}{8},z+\frac{3k_1}{32}+\frac{9k_2}{32}\right) \nonumber \\
    k_4 &= dt.f\left(t+\frac{12dt}{13},z+\frac{1932k_1}{2197}-\frac{7200k_2}{2197}+\frac{7296k_3}{2197}\right) \nonumber \\
    k_5 &= dt.f\left(t+dt , z+\frac{439k_1}{216}-8k_2+\frac{3680k_3}{513}-\frac{845k_4}{4104}\right) \nonumber \\
    k_6 &= dt.f\left(t+\frac{dt}{2} , 
    z-\frac{8k_1}{27}+2k_2-\frac{3544k_3}{2565}+\frac{1859k_4}{4104}-\frac{11k_5}{40}\right)
    \label{rkf2}
\end{align}

We have used the \textit{6th order Runge-Kutta-Fehlberg} technique instead of using the well known \textit{4th order Runge-Kutta} technique to study the transient behaviour near the separatrix. We have taken the time interval $dt = 0.01$ so that the error involved in 6th order RKF method is $dt^6 = 10^{-12}$. This negligibly small error actually helps us to study the transient behaviour very close to the separatrix which wouldn't have been that accurate if 4th order RK method is used where the error involved is $dt^4 = 10^{-8}$

\subsection{Approaching the separatrix on stable manifold}

Can one imagine the separatrix as the critical line observed in equilibrium critical behaviours? Is there any scale of time which shows scale invariance near the separatrix? We are addressing these questions in this study and investigating the transient behaviour near the separatrix, mainly the stable manifold. Keeping the basic idea of basin of attraction in mind, we have solved equation \ref{lv} numerically by varying the initial conditions $(x_0,y_0)$ such that they approach the stable manifold from both right and left. This technique of approaching the separatrix is done for the upper as well as the lower halve of the stable manifold.

\subsubsection{Upper halve of the stable manifold}

The choice of the upper halve has a significant mathematical rigour. The diagram is not bounded by any line(manifold) in the upper halve of the phase diagram in Figure \ref{fig:1}. In the critical phenomenon(like ferro-para phase transition), the para phase is absolutely unbounded. So, this part basically corresponds to the ferro-para phase transition. While approaching the upper halve of the stable manifold, first we have kept the initial value of the population of sheep($y_0$) constant($y_0 = 1.25$) and varied the initial population of rabbits($x_0$). After solving equation \ref{lv}, the population of both rabbits and sheep are plotted with time by varying the initial conditions.

\vskip 0.5cm

\begin{figure*}[htpb]
 \centering
 (a)\includegraphics[width=0.46\columnwidth]{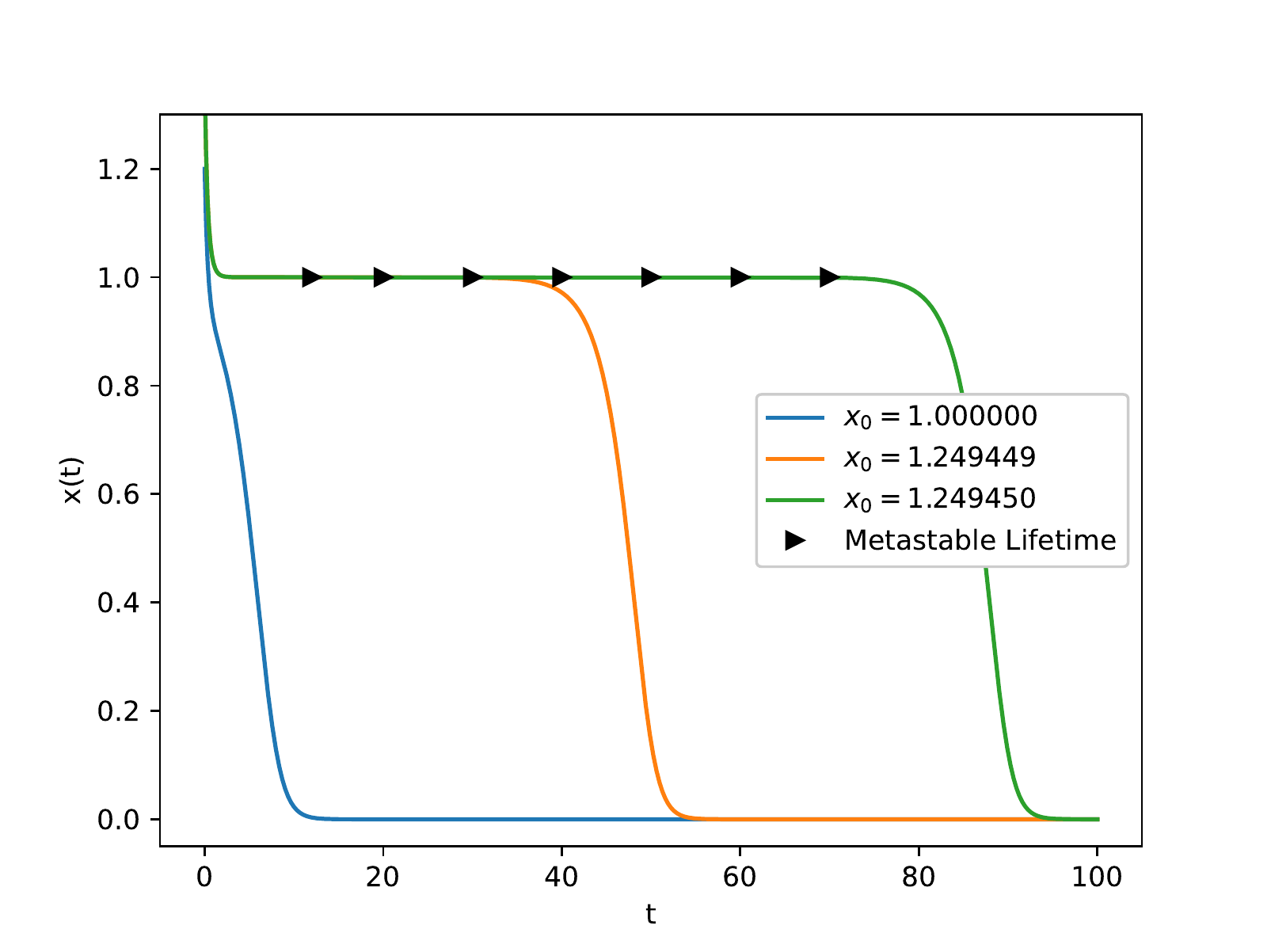}
 (b)\includegraphics[width=0.46\columnwidth]{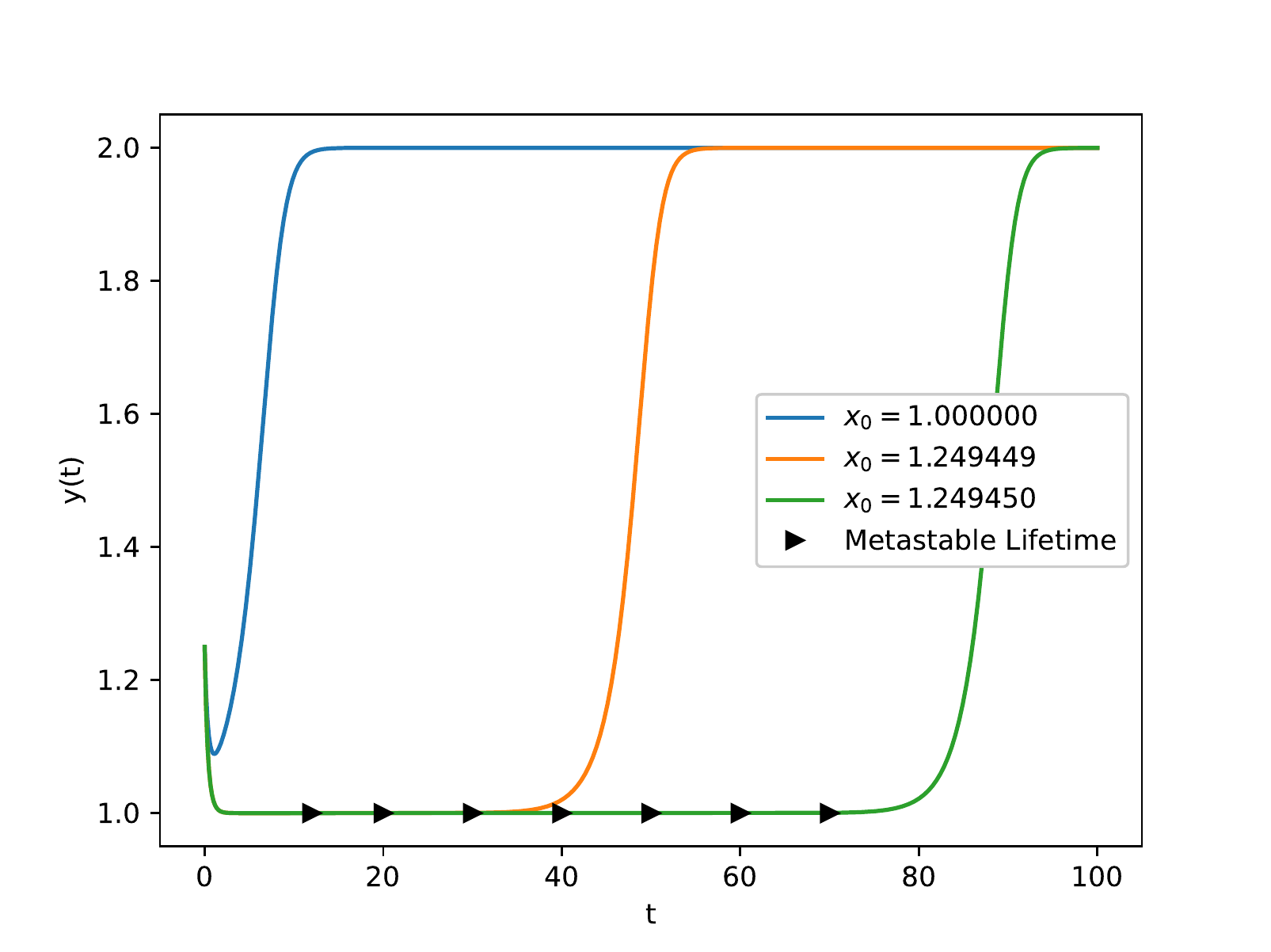}
 \\
 (c)\includegraphics[width=0.46\columnwidth]{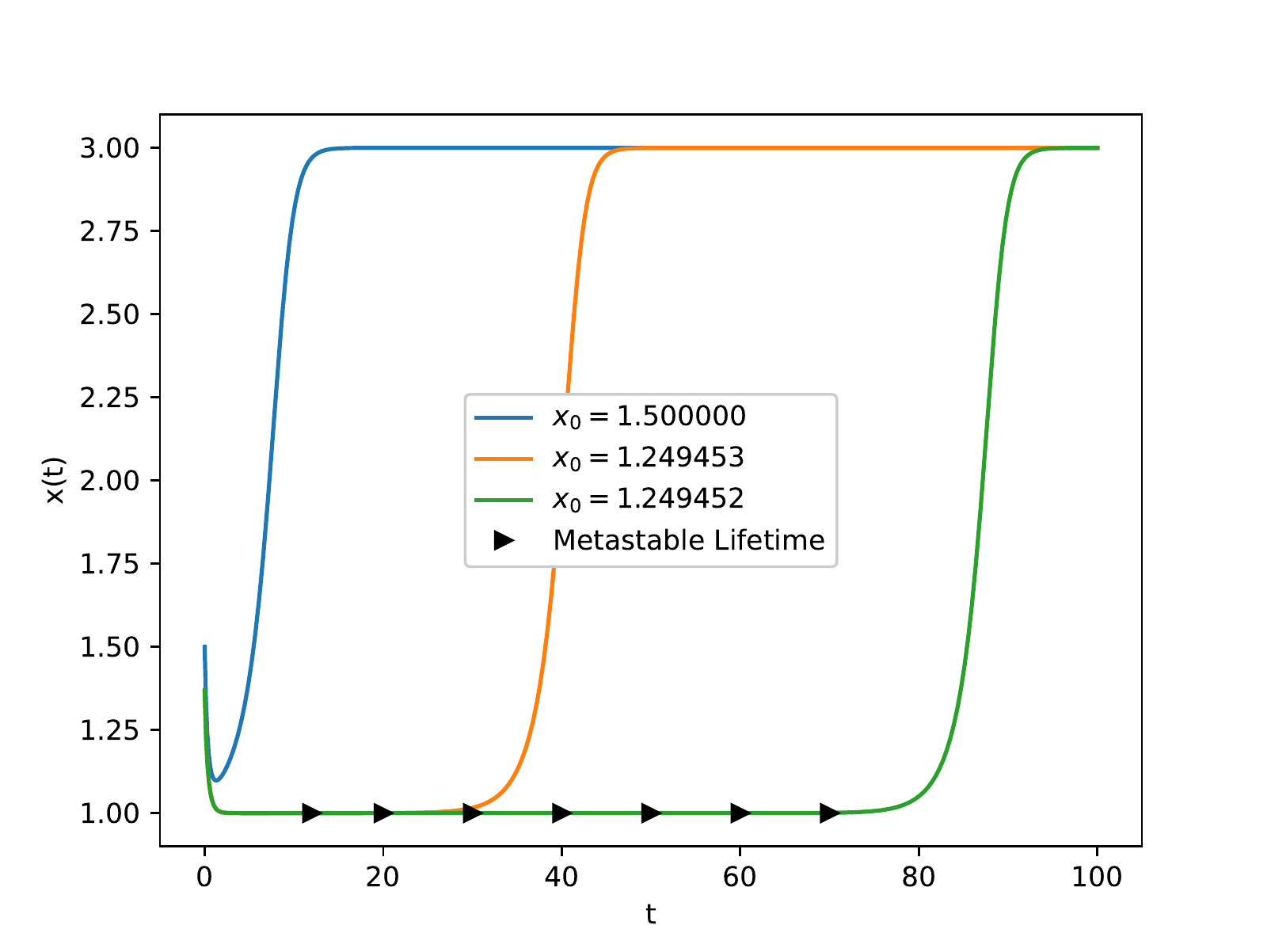}
 (d)\includegraphics[width=0.46\columnwidth]{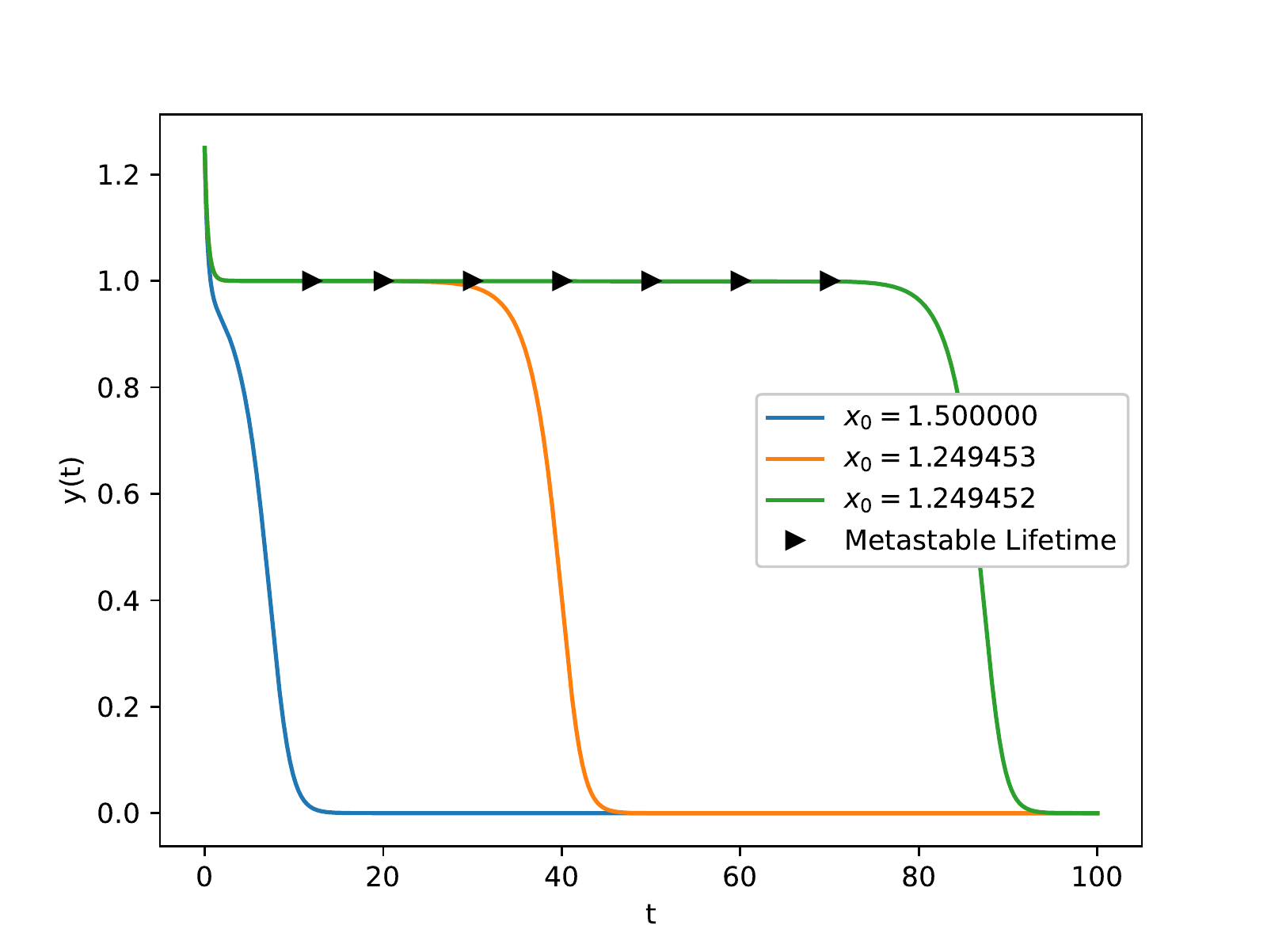}
 \caption{ (Color online) Population as a function of time with initial conditions $x_0$ is changing and $y_0$ (or $h$) fixed at $1.25$. (a),(b) For the left side of stable manifold(Basin for $(0,2)$) where population of rabbits goes to extinction but population of sheep reaches carrying capacity(saturates at $2.0$)). (c),(d) For the left side of stable manifold(Basin for $(3,0)$) where population of rabbits reaches carrying capacity(saturates at $3.0$)) but population of sheep goes to extinction. Spending a finite amount of time at the saddle$(1,1)$ before reaching the final state, indicates the concept of \textit{metastable lifetime}.}
 \label{fig:2}
\end{figure*}

From Figure-\ref{fig:1} we can infer that for the left side of the upper halve of stable manifold(basin for $(0,2)$) as we are approaching closer to a point of the separatrix, we can see that the population of rabbits goes to extinction while the population of sheep is saturated at its carrying capacity. The same behaviour is observed for the right side of the upper halve of stable manifold(basin for $(3,0)$). Here, as we are approaching closer to a point of the separatrix, we can see that the population of sheep goes to extinction while the population of rabbits is saturated at its carrying capacity. This behaviour can be seen in Figure \ref{fig:2}, which was expected from Figure \ref{fig:1}. But we can also observe a significant fact from Figure \ref{fig:2}. Irrespective of which side of the stable manifold we are approaching or irrespective of whether the population is of rabbits or sheep, we can notice that each trajectory spends a finite time at the saddle$(1,1)$ before reaching the final point. This amount of spending time at the saddle increases as we move closer to the separatrix. It has been mentioned in section 2 that $(1,1)$ is a \textit{saddle point}. The time spent in any other state before reaching the ultimate stable state is called \textit{metastable lifetime}. Our observation from Figure \ref{fig:2} signifies that in Lotka-Volterra model, we can see a behaviour equivalent to the concept of \textit{metastable lifetime} in Statistical Mechanics. It is important to note that the metastable lifetime increases as we come closer to the separatrix.

\vskip 0.5cm

\begin{figure*}[htpb]
 \centering
 (a)\includegraphics[width=0.46\columnwidth]{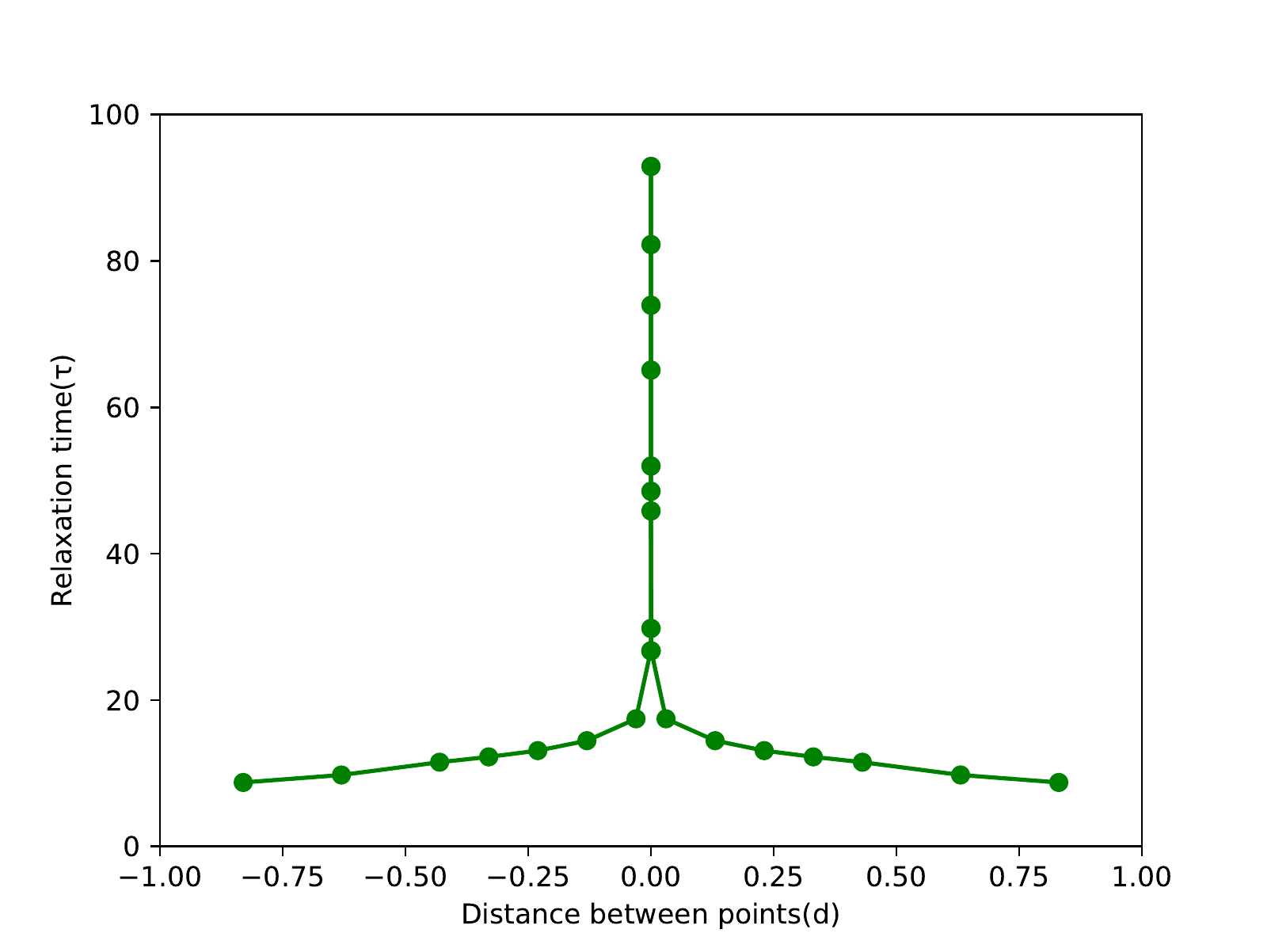}
 (b)\includegraphics[width=0.46\columnwidth]{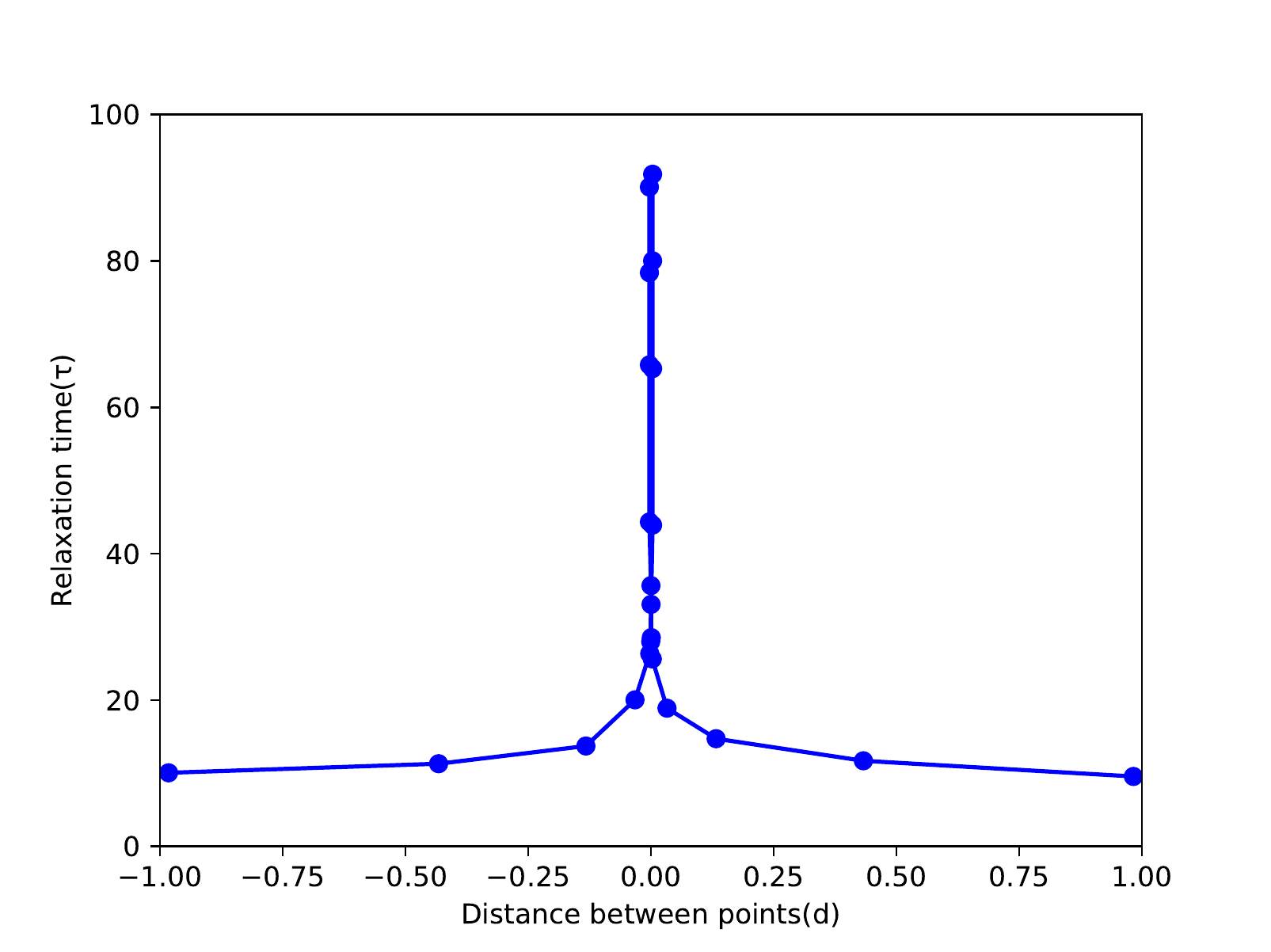}
 \\
 (c)\includegraphics[width=0.46\columnwidth]{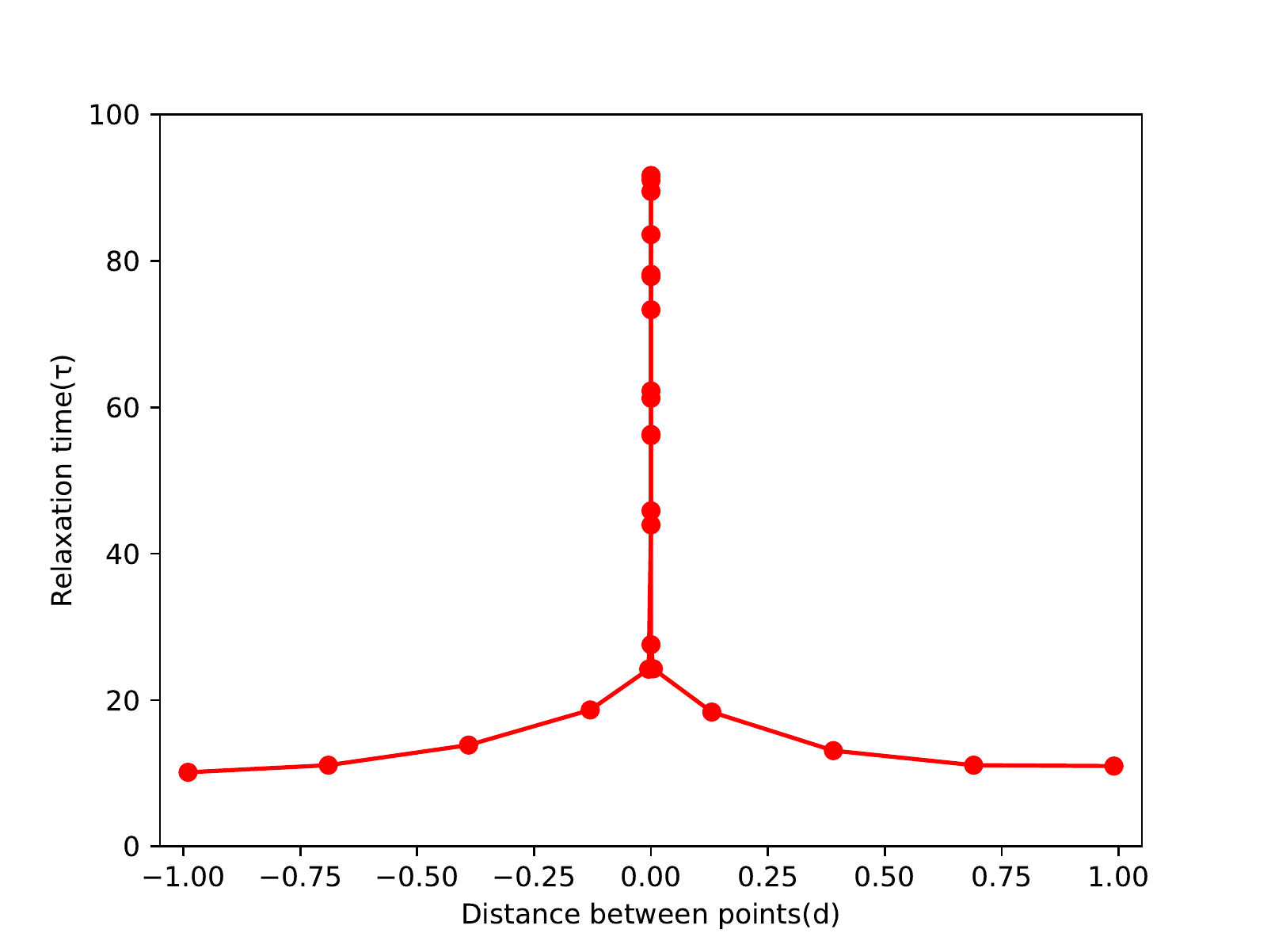}
 
 \caption{ (Color online) Relaxation time($\tau$) as a function of distance($d$) for (a) $h = 1.25$, (b) $h = 1.5$, (c) $h = 1.75$. $\tau$ diverges as $d\to 0$ which indicates the \textit{critical slowing down} near the separatrices in Lotka-Volterra model}
 \label{fig:3}
\end{figure*}

The definition of the basin of attraction is also noticed from Figure \ref{fig:2}. As we come closer to the separatrix(from right or left), the time taken to reach the final state (extinction or carrying capacity) keep on increasing. In other words, very close to any point on the separatrix means very high time taken to reach the final state (stable node). Depending on this observation, we have plotted the distance between points($d$) against the time taken($t_n$) in Figure \ref{fig:3}. The distance between the points is defined as $d = |d_c - d_n|$ where $d_c$ is a constant point on the upper halve of stable manifold and $d_n$ are the points through which we are approaching the separatrix. The plot shows a relaxation behaviour of time taken and that's why rewrite $t_n$ by $\tau$ which is the \textit{relaxation time}. This is plotted in Figure \ref{fig:3} for 3 different values of $y_0$. It is significant to mention that $y_0$ is basically the vertical height($h$) of a point on the stable manifold in the phase portrait.

\vskip 0.5cm

The relaxation time is studied as a function of distance in Figure \ref{fig:3}. It can be noticed that as $d \to 0$ (in other words $d_n \to d_c$), the relaxation becomes slower and when $d \approx 0$ the relaxation time($\tau$) diverges. This is the indication of \textit{critical slowing down}. This fact is demonstrated in Figure \ref{fig:3} for three different heights. We have observed the critically slowing down behaviour from both left as well as right side of $d_c$.

\begin{figure*}[htpb]
 \centering
 (a)\includegraphics[width=0.46\columnwidth]{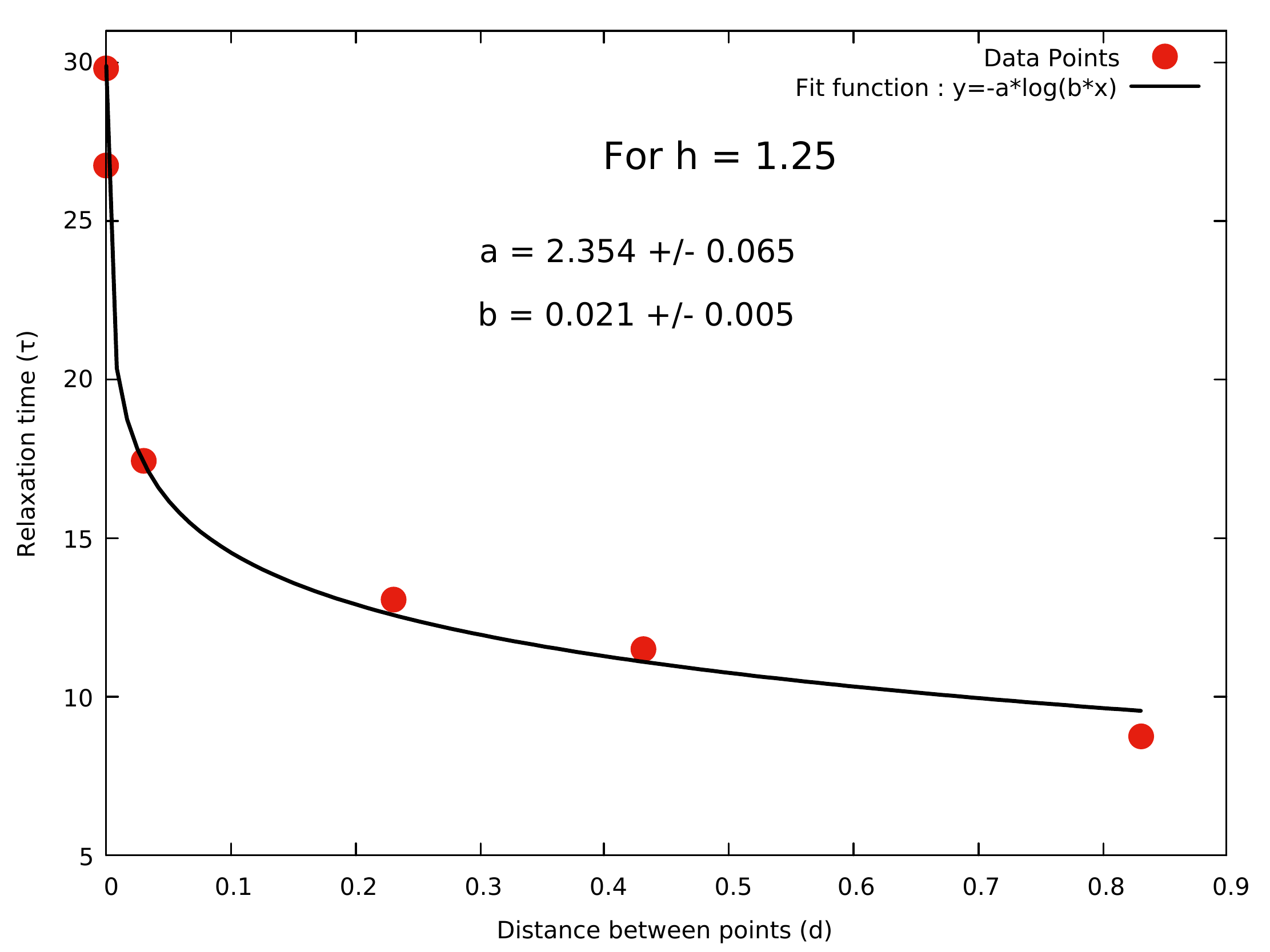}
 (b)\includegraphics[width=0.46\columnwidth]{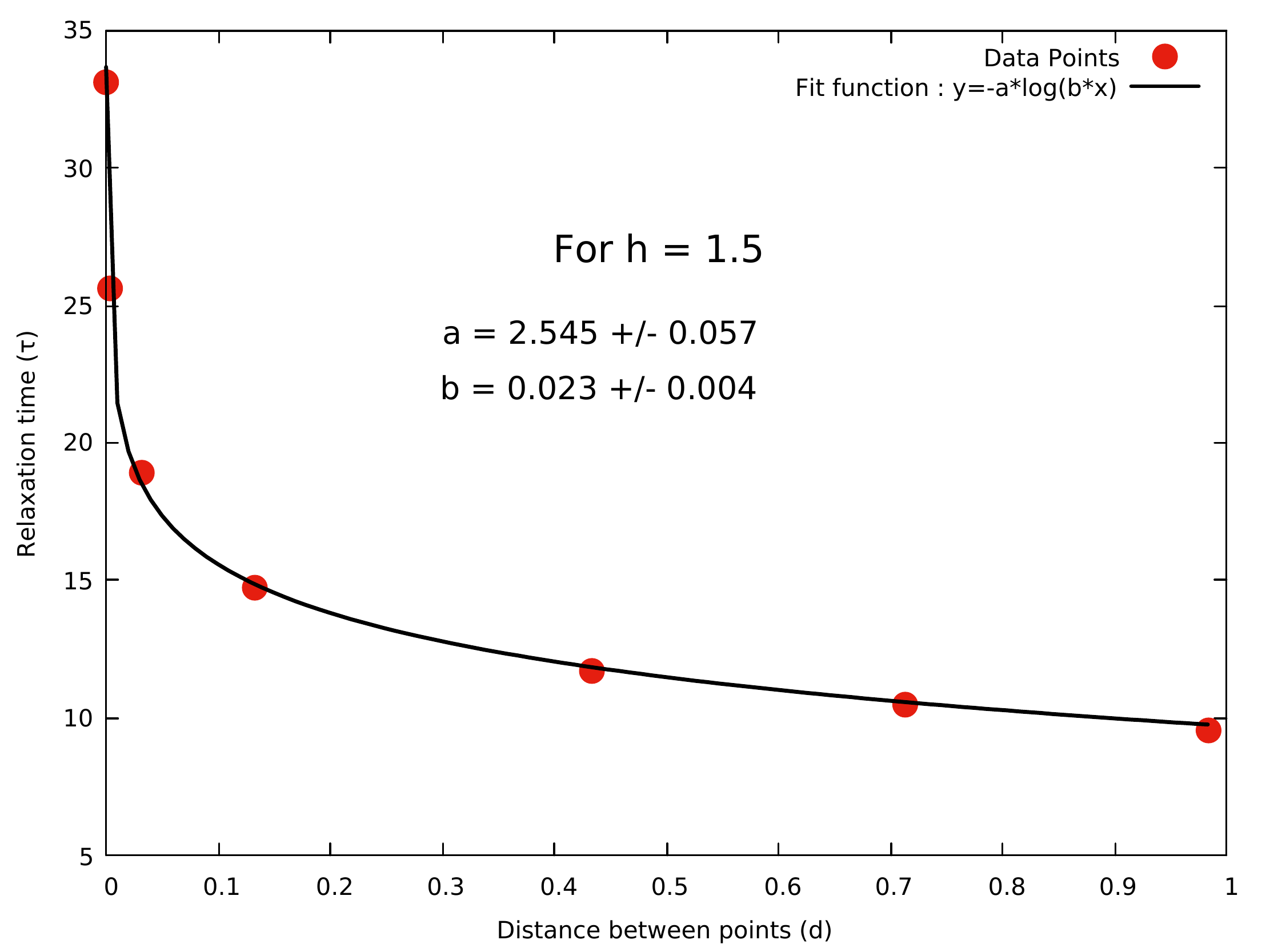}
 \\
 (c)\includegraphics[width=0.46\columnwidth]{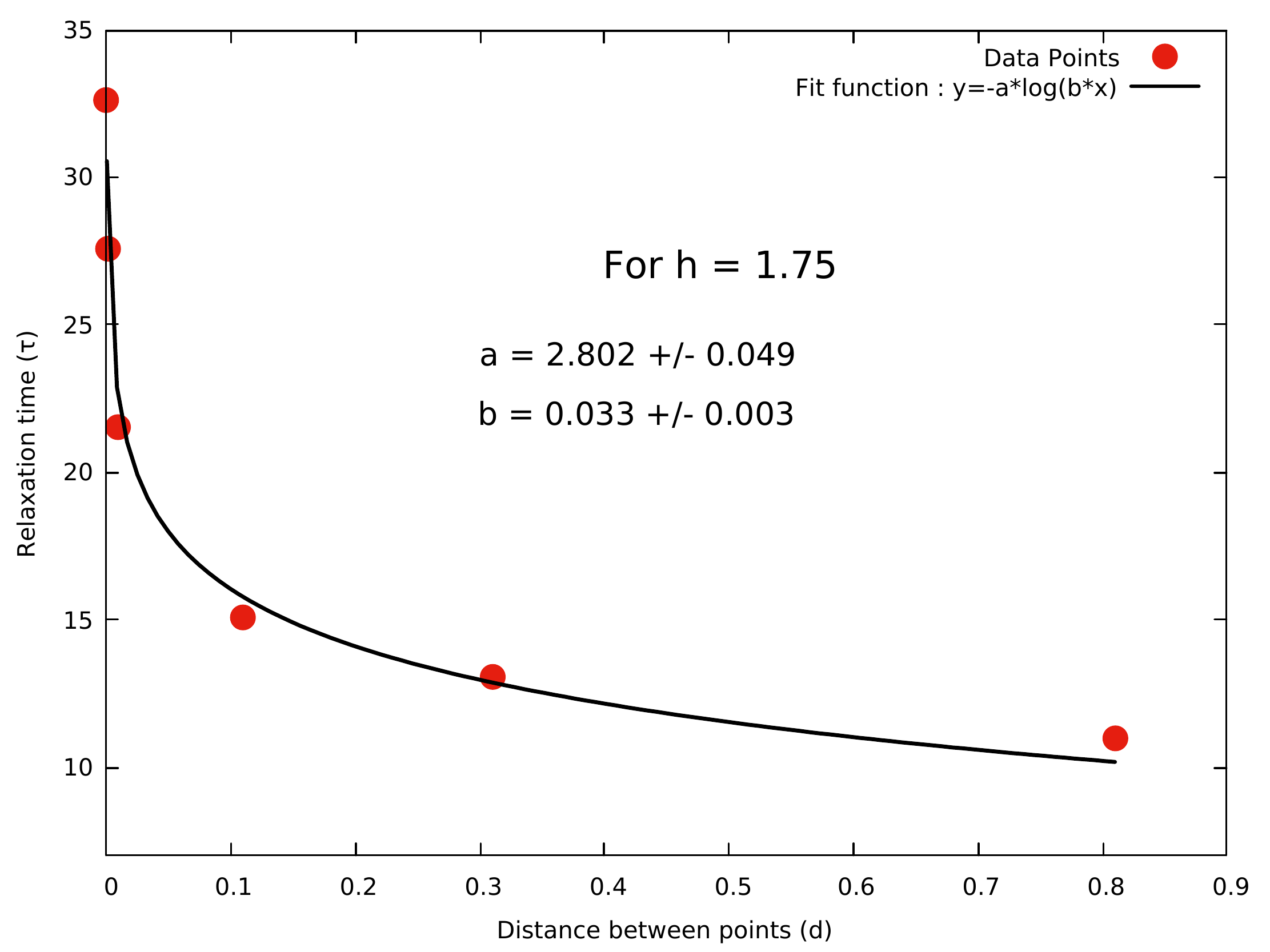}
 
 \caption{ (Color online) $\tau$ is plotted as a function of $d$ for (a) $h = 1.25$, (b) $h = 1.5$, (c) $h = 1.75$. The data points are fitted with a function of the form $y = -alog(bx)$ where (a) $\chi^2 = 0.278$, (b) $\chi^2 = 0.232$, (c)  $\chi^2 = 0.221$. The values of $\chi^2$ confirms the goodness of fit.} 
 \label{fig:4}
\end{figure*}

 We have fitted with a function $\tau \sim -a log(bd)$ (where $a$ and $b$ are fitting parameters)  for three different values of vertical height($h$).  In Figure \ref{fig:4}, we have shown this variation of logarithmic divergence of the time required ($\tau$) to reach the stable fixed point with the distance ($d$) measured from the separatrix. We have considered three different
 heights ($h$) which cut the separatix horizontally. The distance ($d$) is varied on this horizontal line. The fitting parameters ($a$ and $b$) are found to depend on $h$.

\subsubsection{Lower halve of the stable manifold}

\begin{figure*}[h]
 \centering
 \includegraphics[width=0.88\columnwidth]{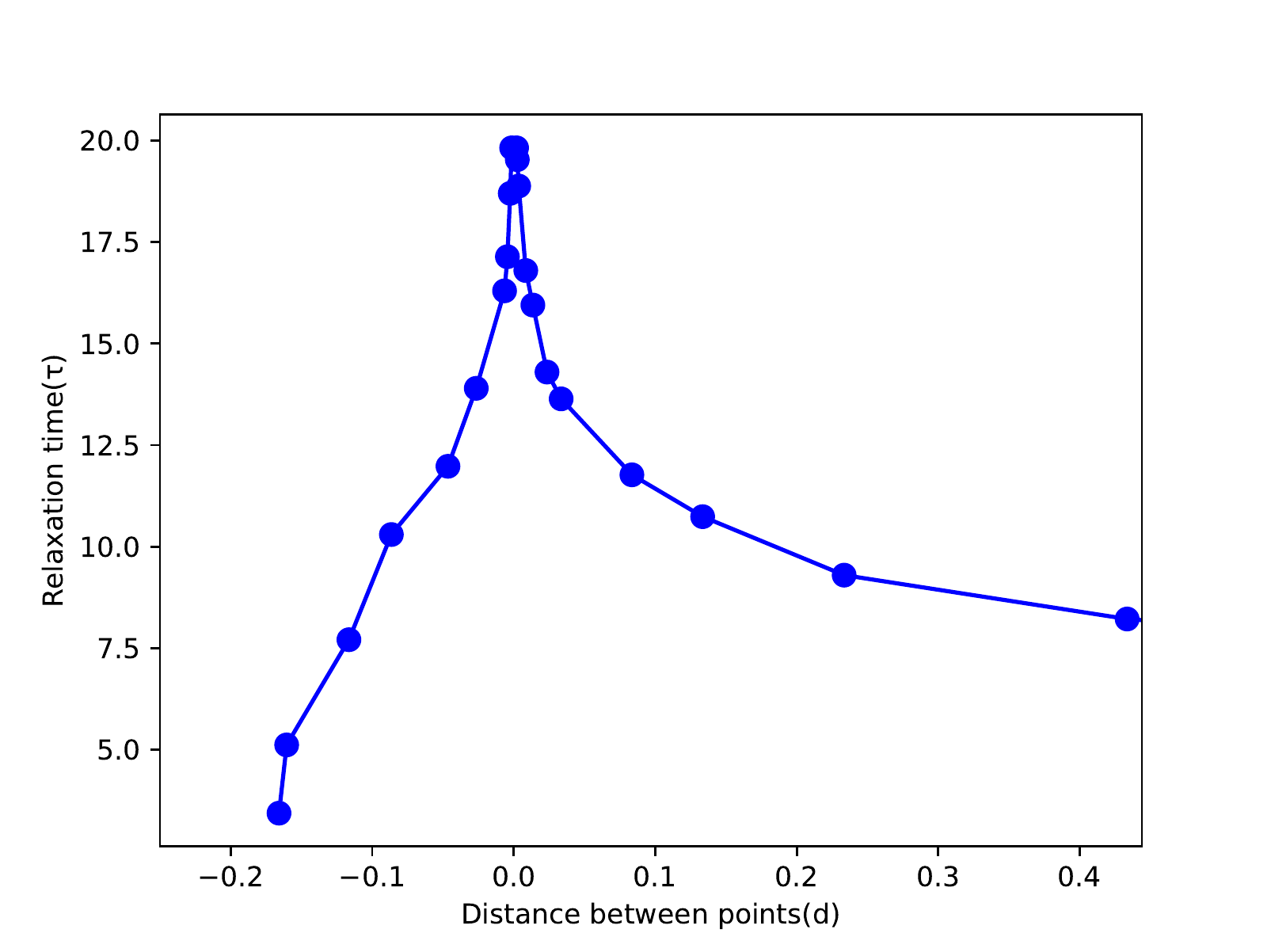}
 \caption{Relaxation time($\tau$) as a function of distance($d$) for the lower halve of stable manifold. Though approaching the separatrix from the right side shows a relaxation behaviour, the relaxation time doesn't actually diverge at $d \to 0$. While approaching the separatrix from the left side shows no relaxation behaviour. Being identical to the ferromagnetic/ordered phase(which is bounded by critical temperature $T_c$), the divergence was also expected here. But the height of the peak is not convincing enough to draw such a conclusion.}
 \label{fig:5}
\end{figure*}

Just like the upper halve, the lower halve also has a physical significance. This region is bounded by lines(manifolds). It corresponds to the ferromagnetic/ordered phase (which is bounded by $T_c$). We have applied the same treatment for the lower halve of the stable manifold. But for this case, the result is a bit different. As we approach the separatrix from right, we see a relaxation like behaviour from the $d$ vs $\tau$ plot, but the relaxation time $\tau$ does not diverge at $d \sim 0$. Whereas, as we approach the separatrix from left side(by changing the initial conditions) we don't find the gentle relaxation behaviour(as was observed previously) from the $d$ vs $\tau$ plot. This is demonstrated in Figure \ref{fig:5}. We can roughly say that the critical slowing down is not exactly observed for the lower halve of the stable manifold. This is due to the boundedness of the lower halve of the stable manifold. It can be seen from Figure \ref{fig:1} that the lower halve of stable manifold is bounded within an area where the population of rabbits and sheep are positive. If we exceed that area(from left side) then we go to second quadrant where one of the population becomes negative which doesn't have a physical meaning. That's why at the left side of the separatrix, we don't see the gentle relaxation. Whereas, for the right side we see a relaxation behaviour because its area is greater than the area of the left side of separatrix as well as in this side there is no value of negative population (due to its location in the first quadrant). But, the relaxation time doesn't diverge here unlike the upper halve of stable manifold because the right side of upper halve was unbounded in the first quadrant whereas the right side of lower halve is bounded by another manifold.   

\vspace{0.3in}
\section{Concluding remarks} 

In this article, we have studied the Lotka-Volterra model of competition (for a fixed amount of food and no predation) by numerical simulation using the Runge-Kutta-Fehlberg algorithm. We have identified the stable fixed points, unstable fixed point, the saddle node, the separatrices of the basins of attraction.

We have tried to know the transient behaviour of this model. Mainly, we tried to know the time ($\tau$) of reaching the stable fixed points from any intial starting point. We have observed that this time ($\tau$) strongly depends on the position of the initial starting point. To characterise it, we have considered the distance ($d$) of the initial starting point measured from the separatrix along a horizontal line in the phase portrait. The time to reach the stable fixed point was found to increase as this initial starting point was closer to the separatix. We have found $\tau \sim -alog(bd)$, i.e., the time $\tau$ diverges
logarithmically (with $d$) as the initial point is chosen closer to the separatix. The separatix is analogous to critical
line (in the case of equilibrium phase transition) and the fact resembles the {\it critical slowing down} with logarithmic divergence.

Such a model lacks spatial variation. So, it is not possible for us to study the divergence of the length scale (correlation length). To study the spatial correlation, the lattice model \cite{provota} should be considered. Another interesting feature of this kind of Lotka-Volterra model is the existence of metastability\cite{stanley}.

If the initial point was chosen below the lower half of the horizontal line passing through the saddle node, the transient behaviour shows a peculiar thing. The stable fixed point has been reached after a metastable lifetime \cite{gunton} closer to the saddle point. It would also be interesting to study various phases and transitions\cite{szabo} in the Lotka-Volterra model. 
\vskip 1cm
\noindent {\bf Acknowledgement:}

SC thankfully acknowledges the library facility of Presidency University, Kolkata. We would like to thank the anonymous reviewer for important suggestions.

\vskip 1cm
\noindent {\bf Authors' contribution:}

{\bf Sauvik Chatterjee} has developed the code for numerical calculations, collected data, prepared figures and written the manuscript.\\

{\bf Muktish Acharyya} has conceptualized the problem, analysed the results and writen the manuscript. 

\vskip 1cm
\noindent {\bf Data availability statement:}
The data will be available on reasonable request to Sauvik Chatterjee.

\vskip 1cm
\noindent {\bf Conflict of interest statement:} We declare that this manuscript is free from any conflict of
interest. The authors have no financial or proprietary interests in any material discussed in this article.

\vskip 1cm
\noindent {\bf Funding statement:} No funding was received particularly to support this work.

\newpage 

\newpage

\newpage

\newpage

\newpage

\end{document}